\documentclass[conference]{IEEEtran}
\IEEEoverridecommandlockouts
\usepackage{amsmath,amssymb,amsfonts}
\usepackage{algorithmic}
\usepackage{graphicx}
\usepackage{textcomp}
\usepackage{xcolor}
\usepackage[backend=bibtex,bibstyle=ieee,citestyle=numeric-comp]{biblatex}
\def \partha [#1]{\textcolor{red}{Partha: #1}} 
\def\BibTeX{{\rm B\kern-.05em{\sc i\kern-.025em b}\kern-.08em
    T\kern-.1667em\lower.7ex\hbox{E}\kern-.125emX}}
\begin{document}

\title{Clotho-AQA: A Crowdsourced Dataset for Audio Question Answering\\
}




\author{
    \IEEEauthorblockN{Samuel Lipping, Parthasaarathy Sudarsanam, Konstantinos Drossos, Tuomas Virtanen}
    \IEEEauthorblockA{\textit{Audio Research Group, Tampere University, Tampere, Finland}}
    \IEEEauthorblockA{\{samuel.lipping, parthasaarathy.ariyakulamsudarsanam, konstantinos.drossos, tuomas.virtanen\}@tuni.fi}
}

\maketitle
\begin{abstract}

Audio question answering (AQA) is a multimodal translation task where a system analyzes an audio signal and a natural language question, to generate a desirable natural language answer. In this paper, we introduce Clotho-AQA, a dataset for Audio question answering consisting of 1991 audio files each between 15 to 30 seconds duration selected from the Clotho dataset. For each audio file, we collect six different questions and corresponding answers by crowdsourcing using Amazon Mechanical Turk. The questions and answers are produced by different annotators. Out of the six questions for each audio, two questions each are designed to have ‘yes' and ‘no' as answers, while the remaining two questions have other single-word answers. For each question, we collect answers from three different annotators. We also present two baseline experiments to describe the usage of our dataset for the AQA task --- a Long short-term memory (LSTM) based multimodal binary classifier for ‘yes' or ‘no' type answers and an LSTM based multimodal multi-class classifier for 828 single-word answers. The binary classifier achieved an accuracy of 62.7\% and the multi-class classifier achieved a top-1 accuracy of 54.2\% and a top-5 accuracy of 93.7\%. 
Clotho-AQA dataset is freely available online at https://zenodo.org/record/6473207.
\end{abstract}

\begin{IEEEkeywords}
audio question answering, Clotho-AQA, dataset
\end{IEEEkeywords}

\section{Introduction}
Question answering (QA) refers to task of providing natural language answers to questions posed in natural language. When a natural signal such as image or audio is used as an auxiliary input, a question can also be targeted to the their contents, leading to visual question answering or audio question answering. The use of natural language enables representing complex high-level information about the inputs such as structure, repetitions and order of events in the case of audio signals. Creating such a multimodal system to answer the question requires inferring information about the contents of the signal that are relevant to the question. Recent advancements in deep learning has made it a suitable choice to tackle these problems. 

Question answering has been largely populated with datasets outside audio with visual question answering~\cite{Malinowski2014AMA,Zhu2016Visual7WGQ,Agrawal2015VQAVQ,Kafle2017VisualQA,Zhu2016Visual7WGQ,Gao2015AreYT}, video question answering~\cite{Lei2018TVQALC,Patel2020RecentAI,Kim2017DeepStoryVS,Yu2019ActivityNetQA}, and textual question answering~\cite{Rajpurkar2016SQuAD1Q,Trischler2017NewsQAAM}. Since most of these datasets contain real-world data (i.e. not automatically generated), they have been annotated by human annotators. Crowdsourcing is a convenient way to do this and has been employed sucessfully in various question answering datasets in modalities outside audio ~\cite{Agrawal2015VQAVQ,Lei2018TVQALC,Rajpurkar2016SQuAD1Q} as well as other multimodal audio-to-text tasks~\cite{drossos2019clotho,kim2019Audiocaps}.

To the authors knowledge, AQA research has seen only two datasets, CLEAR~\cite{abdelnour2018clear} and DAQA~\cite{fayek2020temporal}. CLEAR contains fixed-length audio sequences of different musical notes while DAQA contains generic sound events of variable lengths. The questions and answers in both these datasets are generated programatically. While this allows generating data in a controlled set up, the generated data will most likely lack diversity and challenges that are present in real data. 

Most multimodal QA datasets collect questions and answers from the same annotator ~\cite{Lei2018TVQALC,Gao2015AreYT,Zhu2016Visual7WGQ,Kim2017DeepStoryVS}. Others such as ~\cite{Agrawal2015VQAVQ,Yu2019ActivityNetQA}, use multiple separate annotators for questions and answers. This is useful because different answers might be considered true (e.g. through level of specificity, such as ``bird'' versus ``seagull''). Additionally, separate annotators ensure that only the audio and common knowledge was used to answer the question.

In this paper, we present the first crowdsourced audio question answering dataset, Clotho-AQA. We select $1991$ audio files from the Clotho dataset~\cite{drossos2019clotho} and with our annotation setup, we collect six questions for each audio and three answers for each question using the crowdsourcing platform Amazon Mechanical Turk (AMT). This means that each audio file has an associated $18$ question-answer pairs. Questions and answers are annotated by different annotators. Question annotators are provided with only the audio track and answer annotators are provided with the same audio and the given question. No additional information is given to aid in the annotation. We also present two baseline experiments to show the usage of our dataset. For ‘yes' or ‘no' type questions, we build a multimodal binary ‘yes' or ‘no' classifier and for the other single-word answers we baseline a multi-modal multiclass classifier.

The rest of this paper is divided as follows. Section II introduces the Clotho-AQA dataset, the procedure followed to collect the questions and answers, post-processing steps and data splitting strategy. Section III describes the baseline models, experiments on Clotho-AQA dataset and evaluation metrics. Section IV presents the results of our experiments. Finally, Section V presents our conclusion and future scope.

\section{Creation of Clotho-AQA dataset}
The Clotho-AQA dataset consists of audio files, multiple textual questions related to each file, and textual answers to each question. As audio files, we select $1991$ files from the Clotho dataset at random. The Clotho dataset contains audio files of day-to-day sounds occurring in the environment such as water, nature, birds, noise, rain, city, wind, etc., while avoiding sounds such as music, speech and sound effects.  As such, the audio files are sourced from the Freesound~\cite{freesound} and have a duration between $15$ and $30$ seconds. For details on how the audio files of Clotho were selected and processed, see~\cite{drossos2019clotho}.

\subsection{Question and Answer Gathering}

To collect the questions and answers, we employ a two-step crowdsourcing approach with AMT. For quality assurance, we require that annotators have at least $3000$ approved tasks and an approval rating of $95\%$, while also being located in an English-speaking area (e.g. USA, UK, and CA). We also require question annotators to pass a custom qualification test that includes an English grammar question and a multiple-choice question about the instructions of the question annotation task.

In the first crowdsourcing step, we collect a set of three questions for each audio file. The first question must be answerable with a ‘yes', the second question must be answerable with a ‘no', and the last question must be answerable with a single word that is neither ‘yes' or ‘no'. Secondly, we collect another similar set of three questions. The annotators of the second set of questions are instructed not to ask the same or similar questions to the questions in the first set. In this step, annotators are given an audio track and the previously collected questions. No additional information is provided. After the first step, we then have six questions for each of our 1991 files.

For quality assurance, we manually screen the outputs of the first step by checking the question types (i.e. the ‘yes' or ‘no' questions are not actually single-word questions and vice versa), that the questions are not about contents of speech (e.g. ``Is the language spoken English?''), that the answer does not appear in the question (e.g. ``Does the water flow fast or slow?''), and that the question is not specifically addressed to anyone (e.g. ``Do you want to listen to this?'').

In the second step of data collection, we gather three answers for each of the questions gathered in the first step according to their question type. As additional meta data, we collect confidence scores for the answers from the annotators. For this, we ask the annotators ``Do you feel confident that you were able to answer correctly?''. The annotators are given the options ‘yes', ‘no' and ‘maybe'. In the second step, annotators are given the audio track and one textual question from the first step. No additional information is provided. After the second and final step, we then have $18$ question-answer pairs for each of our $1991$ audio files.

After the data collection, we arrive at a dataset where some of single-word answers are unique to specific audio file. Because we intend to split the dataset such that each audio file only appears in one split (training, validation, or testing), this means that these words will only appear in one split. This leads to sub-optimal training, because trained systems will either learn words unused in validation or will be evaluated on unseen words. For this reason, we count the number of files in which each word appears as an answer. We then employ in-house annotators to replace the answers with a frequency of one. These annotators are instructed to replace these rare answers with other words that appear in the answers of other audio files. This is done by fixing a typographical error, if one is present, or choosing another answer that is as close as possible to the original answer semantically. As a result, the final Clotho-AQA dataset only contains answers with a frequency greater than or equal to two. This post processing reduces the number of unique words from $1889$ to the final dataset which contains a total of $830$ unique words as answers.

\begin{figure}
    \centering
    \includegraphics[width=\linewidth]{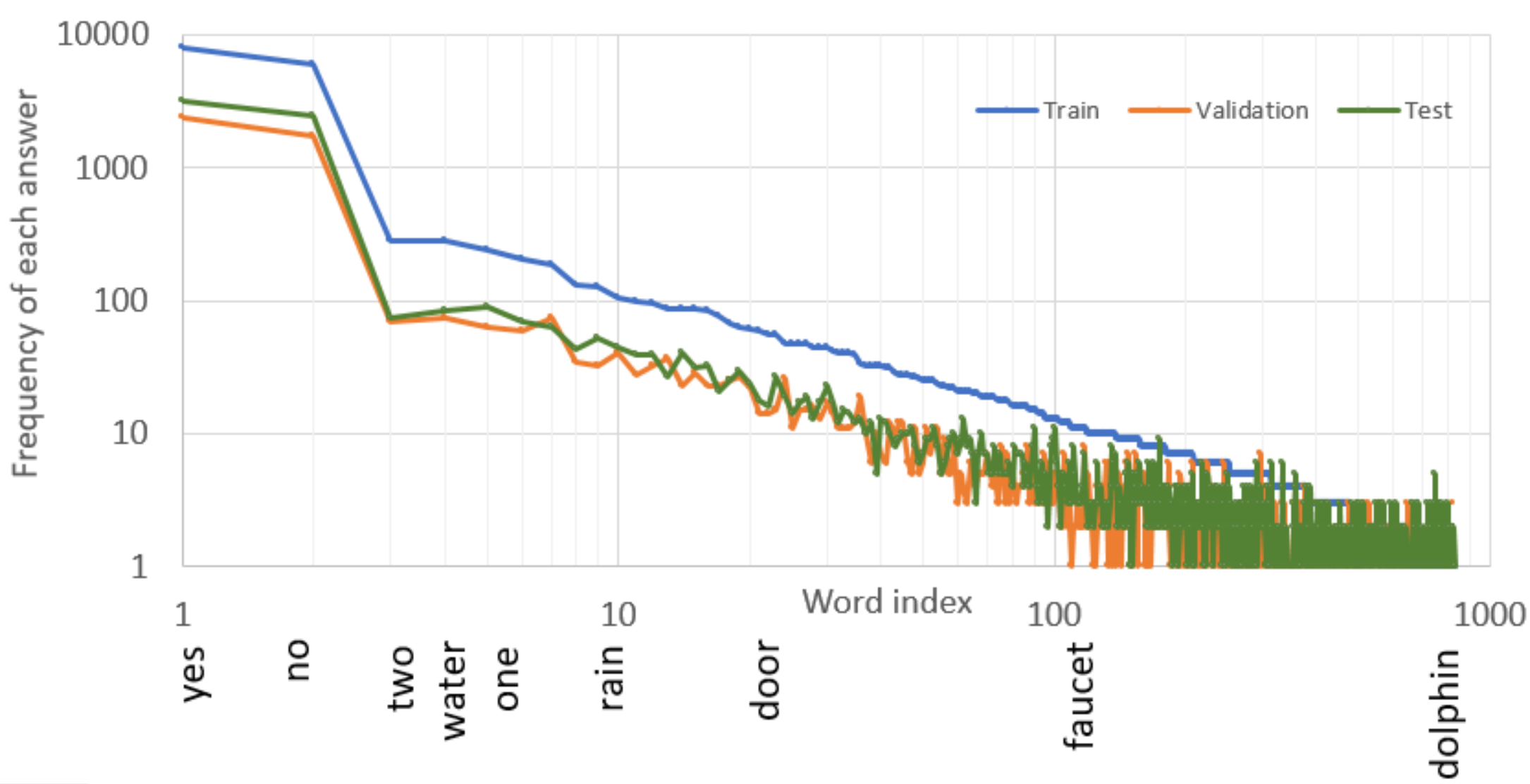}
    \caption{Frequency of words in each of the splits of Clotho-AQA}
    \label{fig:split_word_freqs}
\end{figure}

\subsection{Data Splitting}\label{DS}

We divide the Clotho-AQA dataset into non-overlapping training, validation, and testing splits using the ratio $60\%$-$20\%$-$20\%$. The splits are created such that a given audio file only appears in one split. Additionally, each unique answer should appear in both the training split and one of the validation or testing splits. To this end, we use the same approach as the splitting of the Clotho dataset ~\cite{drossos2019clotho}. For each of the audio files in our dataset we form the answer set of all the unique answer words associated with the audio file. We then use these answer sets as the labels for the audio files and use multilabel-stratification\footnote{\url{https://github.com/trent-b/iterative-stratification}}~\cite{sechidis2011stratification} for split creation.

We create the final splits in two steps. First, we first create $2000$ splits of size $60\%$-$40\%$. We then select the top $50$ splits for further processing, based on an ideal split criteria where each word frequency is divided exactly $60\%-40\%$ between the two splits. Secondly, we split each of the 40\% splits in the above set into half to arrive at splits of size $60\%$-$20\%$-$20\%$. The split that is closest to the ideal split is taken as the final split. For more details of the splitting process, see~\cite{drossos2019clotho}. The final training, validation, and test splits of Clotho-AQA contain $1174$, $344$, and $473$ audio files and $830$, $512$, and $801$ unique answers respectively.

Figure~\ref{fig:split_word_freqs} displays the  frequency of all the unique answers in the final splits.
The two most common answers in the dataset are ‘yes' and ‘no' following the design of the data collection experiment. Single-word answers with a high frequency in the dataset are ‘water', ‘rain', ‘one', ‘two', ‘three', ‘birds' etc.

\section{EXPERIMENTS AND EVALUATION}

To show the usage of Clotho-AQA dataset, we design two baseline experiments and report our initial results. Since ‘yes' or ‘no' type questions constitute two-thirds of our dataset, we create a  binary classifier to evaluate the performance on this subset of the data. We then also design a multi-class classifier for the 828 single-word answers in our dataset.

\subsection{Baseline Methods}\label{AA}

\begin{figure}
    \centering
    \includegraphics[width=\linewidth]{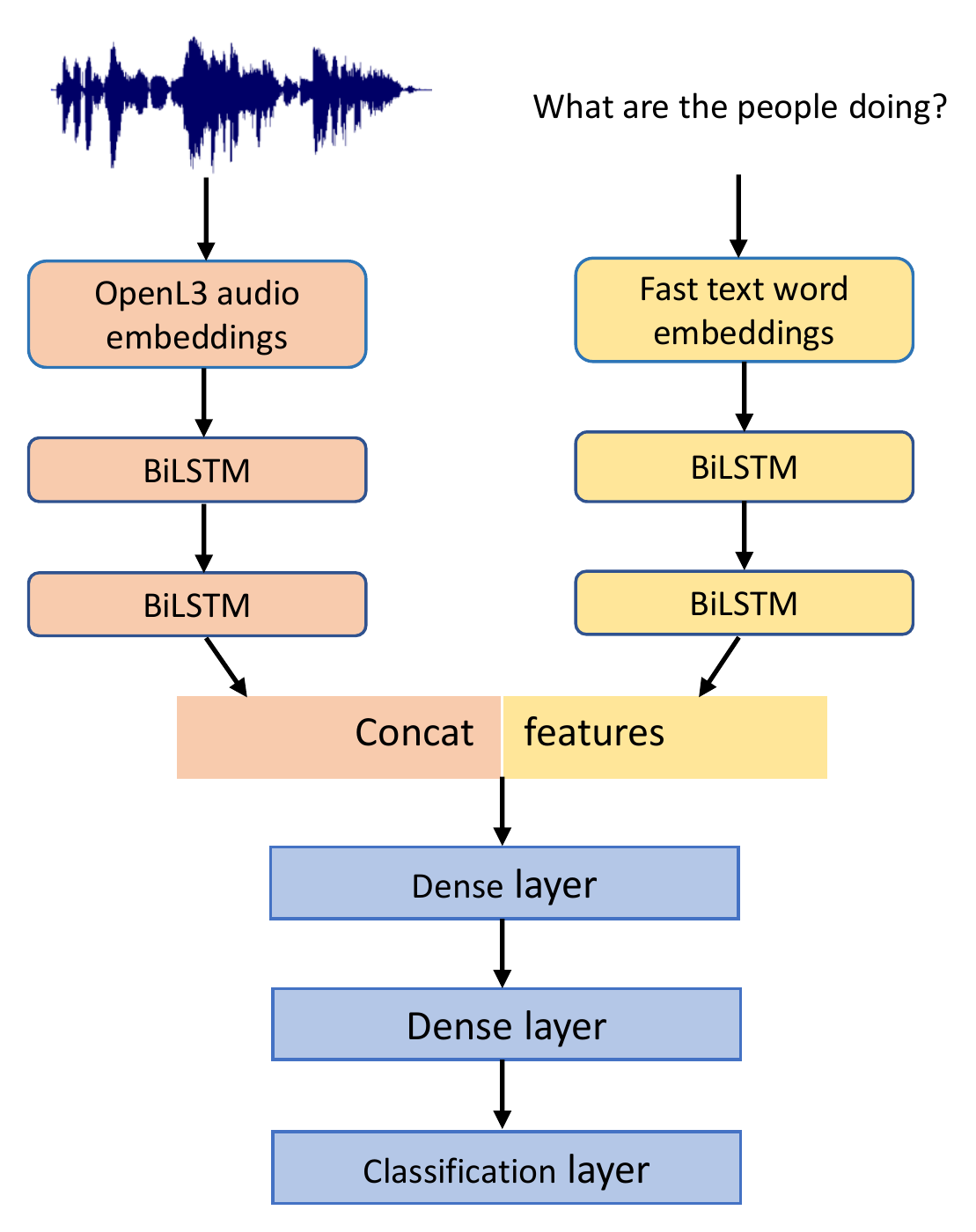}
    \caption{Baseline model architecture}
    \label{fig:model_architecture}
\end{figure}

As the baseline, we use the architecture shown in Figure~\ref{fig:model_architecture}. The inputs to the model are the audio signal and the  question related to the audio. The model contains two branches, one for the input audio and one for the input text. These branches extract features and produce a fixed size representation. The output from these branches are concatenated and passed through fully connected layers for classification. The output from the model is either ‘yes' or ‘no' in case of the binary classification setting or one of 828 single-word answers in case of the multi-class classification setting.

Since the amount of audio samples and textual data in our crowdsourced dataset are relatively small to learn a good representation, we use pre-trained embeddings to extract features. For the input audio signal, we use OpenL3 ~\cite{8682475} to compute deep audio embeddings. OpenL3 is based on L\textsuperscript{3}-Net~\cite{DBLP:journals/corr/ArandjelovicZ17} trained on Audioset~\cite{45857} videos in a self-supervised setting for audio-visual correspondence task. OpenL3 provides two models based on the content type --- one trained on musical data and one on environmental data. Since most of our audio signals are common environmental sounds, we chose the latter model. The input to this model is $\mathbf{X}\in\mathbb{R}^{T\times 128}$ mel-spectrogram of the audio with 128 mel bands and $T$ time frames. The output from the OpenL3 pre-trained embeddings is $\mathbf{X}_{emb}\in\mathbb{R}^{T'\times 512}$,  where $T'$ is the number of output time frames from the OpenL3 model and 512 is the audio embedding size. An hop size of 0.1 seconds is used to extract the embeddings. The audio embeddings are then passed through a series of bidirectional LSTM layers, $\text{BiLSTM}_{n}$ with $n = 1, 2$ to learn temporal relationships and to convert them into a fixed size representation. The  bidirectional LSTM is given by

\begin{equation}
    \mathbf{X}_{n} = \text{BiLSTM}_{n}(\mathbf{X}_{n-1}),
\end{equation}

\noindent
where $\mathbf{{X}}_{0}$ = $\mathbf{{X}}_{emb}$. If $h$ is the number of hidden units in the BiLSTM, then $\mathbf{X}_{n}\in\mathbb{R}^{T'\times 2h}$. We then choose as output $\mathbf{{x}}_{n}\in\mathbb{R}^{2h}$, the final time step of the last BiLSTM layer to represent the fixed size audio embedding.

For the natural language question input, we represent the question as word embeddings. The embedding for each word is computed using the Fasttext \cite{mikolov2018advances} pre-trained word vectors. The Fasttext word vectors we use have 1 million word vectors trained on Wikipedia 2017, UMBC webbase corpus and statmt.org news dataset. If the input question $\mathbf{Q}$ has $K$ words, the word embeddings using Fasttext is $\mathbf{Q}_{emb}\in\mathbb{R}^{K\times 300}$, where 300 is the size of the word embedding. The word embeddings are also passed through a series two of bidirectional LSTM layers. If $h'$ is the size of the hidden units in the BiLSTM, 
we again choose as output $\mathbf{{q}}_{n}\in\mathbb{R}^{2h'}$, the output of the final time step in the last BiLSTM layer to represent the fixed size word embedding for the input question. For the binary classifier, we use a hidden size of 128 for the bidirectional LSTM layers with a dropout of 0.2 for both the audio and text branches. In case of the multiclass single-word answers classifier, we use a hidden size of 512 for both the branches. These hyper parameters were tuned based on the model's performance on the validation split.

The outputs from both the audio and the question branches are concatenated and then passed through a series of fully connected layers $\text{Dense}_{k}$ with $k = 1, 2$ with ReLU non-linearity. The fully connected layers combine the learnt features of both the the audio and the textual question. This is given by  

\begin{equation}
    \mathbf{D}_{k} = \text{Dense}_{k}(\mathbf{D}_{k-1}), 
\end{equation}

\noindent
where $\text{D}_{0} = concat[\mathbf{{x}}_{n}, \mathbf{{q}}_{n}]$, and \textit{concat} represents the concatenation operator. The two dense layers in the binary classifier have 256 and 128 neurons respectively, while they have 1024 neurons each in case of the multiclass classifier. The bigger size in case of the multiclass classifier increases the model's capacity to capture finer details to classify among 828 different answer classes.

The output of the final dense layer is then passed to a classification layer, which is again a dense layer with the number of neurons equal to the number of unique answer  classes. The output of the classification layer is $\hat{y}\in\mathbb{R}^{C}$, where $C$ is the number of answer classes in the dataset. The classification layer is a simple logistic regressor with one neuron in case of the ‘yes' or ‘no' binary classifier.  In case of the multiclass classifier, the final classification layer contains 828 neurons, one for each unique single-word answers followed by a softmax activation.

\subsection{Evaluation}
We trained and evaluated the baseline models on the Clotho-AQA data splits obtained as described in Section \ref{DS}. The dataset contains 18 question answer pairs for each file. To create the splits for the binary classifier, we select the ‘yes' or ‘no' questions from the respective data splits. We end up with the same number of audio files 1174, 344, 473 for training, validation and testing respectively with each file having 12 ‘yes' or ‘no' question-answer pairs. Similarly, for the multi-class classification task, we select the single-word answers from the respective data splits resulting in the same number of audio files as the original splits with each file having six question answer pairs.


To analyze the performance of the binary classifier on contradicting answers to the same question by different annotators, we also train and evaluate the binary classifier on three different subsets of the data. The first case is where we consider all the question-answer pairs as valid inputs even if they have contradicting answers. Secondly, we train and evaluate the model on only those question answer pairs where all the three annotators have responded unanimously. Finally, we create a subset using majority voting strategy where, for each question, we choose as the label, the answer provided by at least two out of the three annotators. These three cases are denoted as ‘Unfiltered data', ‘Unanimous' and ‘Majority votes' respectively in Table \ref{tab:binary_results}.

Further, we also train models with only one of the multimodal inputs, i.e, a model with only the textual question as input with no auxiliary audio input and a model with only the audio signal as input. This helps us to analyze how well the model captures the information from both the modalities in predicting the answer.

All the models are trained with Adam optimizer with a learning rate of 0.001, $\beta_1 = 0.9$, $\beta_2 =0.999$. The models are trained for 100 epochs with cross-entropy loss and the model with he best validation score is used for testing.

\section{Results}
Table \ref{tab:binary_results} summarizes the results of all our experiments on Clotho-AQA dataset for binary classification of ‘yes' or ‘no' questions. It is clear from the results that the model performs better when the answers are unanimous indicating intelligible presence of the answer in the audio compared to the case where annotators give different answers to the same question. It is interesting to note that the model that takes in only the question performs as good as the model that takes both the inputs. This behaviour has also been commonly observed in many VQA taks ~\cite{VQA_err1, VQA_err2, VQA_err3}. This could be due to strong priors in the language model, our shallow baseline network or poor audio embeddings extracted from the pre-trained model for our dataset. The results of our single-word multiclass classifier are presented in Table \ref{tab:swc_results}. Since the number of unique answer classes are high (828), we also use top-5 accuracy and top-10 accuracy  metrics to evaluate the performance of the classifier. The results indicate that the model is starting to learn the relationships between the multimodal data. Here again, the model with only the question input performs as good as the model with multimodal input. In future, we plan to analyze the reasons for such behaviour and improve the data and model. 

\begin{table}[htbp]
\caption{Accuracies (\%) of binary ‘yes' or ‘no' classifier on Clotho-AQA}
\begin{center}
\begin{tabular}{|c|c|c|c|}
\hline
\textbf{Input}&\textbf{Unfiltered data} &\textbf{Unanimous} &\textbf{Majority votes} \\
\hline
Audio only & $57.5$ & $62.1$ & $58.2$\\
\hline
Question only & $63.5$ & $71.8$ & $64.4$ \\
\hline
Audio + question & $62.7$ & $73.1$ & $63.2$\\
\hline
\end{tabular}
\label{tab:binary_results}
\end{center}
\end{table}

\begin{table}[htbp]
\caption{Accuracies (\%) of single-word answers multi class classifier on Clotho-AQA}
\begin{center}
\begin{tabular}{|c|c|c|c|}
\hline
\textbf{Input}&\textbf{Top-1 } &\textbf{Top-5 } &\textbf{Top-10} \\
 &\textbf{ accuracy} &\textbf{ accuracy} &\textbf{ accuracy} \\
\hline
Audio only & $3.2$ &$13.4$ & $21.1$\\
\hline
Question only & $55.7$ & $96.8$ & $99.4$\\
\hline
Audio + question & $54.2$ & $93.7$ & $98.0$\\
\hline
\end{tabular}
\label{tab:swc_results}
\end{center}
\end{table}

\section{Conclusion}

In this paper, we present Clotho-AQA, an audio question answering dataset consisting of 1991 audio files selected from Clotho dataset. For each audio file, six questions were collected from native English speaking countries by crowdsourcing using AMT. The questions are collected such that it is possible to answer the question with a single-word or ‘yes' or ‘no'. The answers are also collected by crowdsourcing using AMT. For each question, the answers are collected from three different annotators. We then post-process the data to remove unique words and replace with commonly occurring suitable answers. We also present baseline models to show how to use the dataset. We trained a binary classifier for ‘yes' and ‘no' answers and a multi-class classifier for single-word answers. Currently, the question only model produces equally good results as the multimodal input model. Going ahead, we intend to analyze these results and improve the data and the model. In future, the Clotho-AQA dataset can be used for development of novel methods for the audio question answering tasks.

\section*{Acknowledgment}

The authors wish to acknowledge CSC-IT Center for Science, Finland, for computational resources used towards this research.

 \printbibliography

\end{document}